\documentclass{ws-mpla}
\usepackage[super]{cite}
\usepackage{xcolor}
\usepackage[verbose,hypertexnames=false]{hyperref}
\usepackage{float}
\usepackage{subfig}
\usepackage{hyperref}
\usepackage{color}
\usepackage[english]{babel}
\usepackage[utf8]{inputenc}
\usepackage{comment}
\usepackage{cleveref}
\usepackage{bm}
\usepackage{braket}
\usepackage{url}

 \newcommand{\eqn}[1]{Eq.\,(\ref{#1})}
 \newcommand{\secref}[1]{section \,(\ref{#1})}

\newcommand{\orderof}[1]{ \mathcal{O} \left( #1 \right) }

\newcommand{\ee}{\end{eqnarray}}

\newcommand{\ave}[1]{\left\langle #1 \right\rangle}
\newcommand{\eqcomma}{\phantom{AA},\phantom{AA}}

\begin{document}
\begin{center}
{\Large 
\textbf{The quantum Newton's bucket: Active and passive rotations in quantum theory}
}
\end{center}

\begin{center}
Augusto Facundes, Kayman J. Gonçalves, Giorgio Torrieri
\end{center}

\begin{center}
Institute of Physics ``Gleb Wataghin'', University of Campinas, Campinas, Brazil
\end{center}
Motivated both by classical physics problems associated with ``Newton's bucket'' and recent developments related to QCD in rotating frames of reference relevant to heavy ion collisions, we discuss the difference between ``active'' and ``passive'' rotations in quantum systems.  We examine some relevant potentials and give general symmetry arguments to give criteria where such rotations give the same results.   We close with a discussion of how this can be translated to problems of current interest in quantum field theory and quantum gravity.
    \section{Introduction \label{intro}}
    Isaac Newton, in his work ``Philosophiæ Naturalis Principia Mathematica" \cite{newton,assis}, discusses a thought experiment in which a cylindrical container, such as a bucket, is partially filled with a fluid, like water, and is suspended in the air by a rope. After twisting the rope sufficiently and releasing it, the bucket rotates with a constant angular velocity. At the beginning of the motion, only the bucket rotates, while the water remains at rest. Then, the water gradually starts rotating with the same angular velocity as the bucket, forming a surface in the liquid that takes the shape of a paraboloid of revolution—different from when the water is at rest, where the surface is flat. This curvature remains for some time after the bucket stops rotating, returning to a flat surface when the water also comes to rest. The thought experiment demonstrates that the physical effects of rotation are absolute, not relative. Examining each stage in detail:
    \begin{description}
        \item[Initial rotation of the bucket] bucket in rotation, water at rest, water not curved;
        \item[Rotation of the bucket in progress] bucket in rotation, water in rotation, water curved;
        \item[Bucket stops rotating] bucket at rest, water in rotation, water curved;
        \item[Water stops rotating] bucket at rest, water at rest, water not curved.
    \end{description}
    
    It is clear that the physical effect of the water's rotation (the curvature) is not sensitive to the water's rotation relative to the bucket but rather to the absolute rotation of the water. An observer moving in circles around the bucket would easily, by observing the lack of curvature in the water, ascertain that it is he rather than the water which is rotating.
    
    This is quite different from the ``relativity of velocity" in Galileo's principle of relativity and justifies Newton's claim that space is, in some sense, absolute. Mach attempted to define ``absolute acceleration" in terms of the asymptotic boundary (``distant stars"), whereas Einstein postulated that acceleration is defined in terms of local spacetime, dynamically determined by gravity. The relationship between the two definitions, and the implications for dynamics, gravity and cosmology are still subject to debate in the literature \cite{rothman}.
    
    We were reminded of these classic conudrums by the recent interest in the literature of ``QCD in rotating referece frames'' \cite{rot1,rot2,rot3,rot4,rot5,rot6,rot7} motivated by the experimental finding of vortical effects in Quark-Gluon plasma \cite{lisabec,voloshin,qun}.   A variety of both perturbative and non-perturbative approaches have been used to calculate quantities in a rotating medium.   Yet the way rotation was imposed was essentially to work with a metric put in by hand, which corresponds to a passive rotation.   Of course, in a laboratory, systems rotate because there is something in their dynamics that makes them rotate.  To what extent are the two equivalent?  
    Some of the works in \cite{rot1,rot2,rot3,rot4,rot5,rot6,rot7} use the ``rigid body`` approximation (where, for conserved angular momentum, active and passive rotations are indeed trivially equivalent), but since rigidity is violated both by relativity and quantum mechanics, there is no clear sense this is a well-defined effective theory.
    
    In the rest of this work we shall investigate this ``elementary'', but surprisingly subtle question in a variety of systems, and will try to establish criteria for this equivalence.
     
    One has to be clear as to what ``active'' and ``passive'' mean mathematically, especially as the concept of accelleration in general only makes sense in the semi-classical limit (the Newton relation $d\hat{p}/dt=m dV/d\hat{x}$ becomes ill-defined if commutators between position and momentum are non-negligible).
    
    An active rotation is one where the rotation is due to the Hamiltonian of the system, so 
    \begin{equation}
    \label{activedef}
        \hat{H} \rightarrow \hat{H}+\hat{H}_{\text{rotation}} \eqcomma |\psi\rangle \rightarrow \exp[i \hat{H}_{\text{rotation}} t] \ket{\psi}
    \end{equation}
    or equivalently, given a complete set of Eigenstates $\ket{\psi}_i$ and coefficients normalized $c_{ij}$
    \begin{equation}
        \hat{\rho} = c_{ij} \ket{\psi_i}\bra{\psi_j}\eqcomma   \frac{d\hat{\rho}}{dt} = i[\hat{\rho}, \hat{H}]
            \label{eq:heisenberg}
        \end{equation}        
        where the classical counterpart of the Hamiltonian $\hat{H}$ includes a rotation potential and the state of the system is close to a ``wavepacket''.
        
    In a passive rotation, the system is stationary, and the detector rotates around the system.   This can be accomplished by
    \begin{equation}
    \label{passivedef}
    \hat{\rho} \rightarrow U^\dagger(t) \hat{\rho} U(t),
    \end{equation}
    where $U(t)$ is a rotation matrix whose angle is $\sim \omega t$.
    This argument leads to a relatively straight-forward conclusion.  If the active rotation potential is equivalent to shifting the Hamiltonian 
        \begin{equation}
        \label{criterion}
            \hat{H} \rightarrow \hat{H}_0 - \boldsymbol{\omega}\cdot\mathbf{\hat{J}} \eqcomma [\mathbf{\hat{J}},\hat{H}_0]=0,
        \end{equation} 
        both rotations are the same, because the ``passive" rotation matrix is exactly the same as the difference between the Hamiltonian evolutions (``before" and ``after" the shift).   As we shall see,  a crucial requirement is that $[\mathbf{J},H_0]=0$;  This requirement is ``quintessentially`` quantum, as it reflect the relationship between lack of commutation and experimental uncertainity.

In the following sections, we will examine in detail a series of examples related to textbook quantum mechanics, computing in detail passive and active rotations and comparing/contrasting the results.  We then extract general principles we can use to understand the difference between passive and active rotations, and finish by discussing the lessons that can be drawn for problems at the frontier of physics research, including quantum field theory and combining gravity with quantum mechanics.
        \section{Quantum mechanics \label{quantum}}
        In the following,   we consider a ``high-lying'' energy state where one can think of ``a rotating wavepacket'' rather than a wavefunction, under a variety of potentials.   Physically, this system is close to what is known as a Rydberg atom \cite{rydberg}
    \subsection{Coulombic potential \label{coloumb}}
    Let us consider a particle with a generic spin in the presence, at first, for simplification, of a Coulomb-like scalar potential $V(r) = -\alpha/r,\, \alpha \in \mathbb{R}$.  

    \subsubsection{Active rotation}
        In the reference of center of mass, the particle under rotation is subjected to the action of the Hamiltonian
        \begin{equation}
        \hat{H} = \frac{\mathbf{\hat{p}}^2}{2m} - \frac{\alpha}{r} - \boldsymbol{\omega}\cdot\mathbf{\hat{J}},
            \label{eq:hamiltoniana-ativa}
            \end{equation}
        where $\omega$ is the angular velocity of oscilation and $\mathbf{\hat{J}}$ is the total angular momentum operator. 
         The density operator at the initial instant ($t = 0$) has the form
        \begin{equation}
            \hat{\rho}(0) = \rho_{m_1, m_2}|r\rangle\langle r'|\otimes  |S\rangle\langle S'|,
            \label{eq:rho_0}
        \end{equation}
        where $|r\rangle = |n, \ell, m_{\ell}\rangle$ represents the states in the position space, $|S\rangle = |s, m_s\rangle$ the states in the spin space and $\rho_{m_1, m_2}$ the normalization coefficients of the density matrix at $t = 0$. 
        Rewriting the tensor product in (\ref{eq:rho_0}) in the form
        \begin{equation}
            \hat{\rho}(0) = \rho_{m_1, m_2}|n, \ell, m_\ell; s, m_1\rangle\langle n, \ell, m_\ell; s, m_2| 
            \label{eq:rho_0-compressed}
        \end{equation}
        and defining $J = \ell + s$ the total angular momentum and $M = m_\ell + m_s$ the total magnetic momentum, the state $|n, J, M\rangle$ can be writen as
        \begin{equation*}
            \displaystyle |n, J, M\rangle = \sum_{m_\ell, m_s} |n, \ell, m_\ell; s, m_1\rangle\langle n, \ell, m_\ell; s, m_2|n, J, M\rangle
        \end{equation*}
        where the factors $\langle n, \ell, m_\ell; s, m_2|J, M\rangle$ inside the sum define the set of Clebsch-Gordan coefficients ($C^{J, M}_{s, m_\ell, \ell, m_2}$) and, therefore,
        \begin{equation}
            \displaystyle |n, J, M\rangle = \sum_{m_\ell, m_s} C^{J, M}_{s, m_\ell, \ell, m_2} |n, \ell, m_\ell; s, m_1\rangle
            \label{eq:n,J,M}
        \end{equation}
        represents the state in which the angular and magnetic momenta of the particle under (active) rotation couple.
        
        From (\ref{eq:n,J,M}), the elements of $\hat{\rho}(0)$ are given by
        \begin{equation*}
            \displaystyle\langle n, J, M_1|\hat{\rho}(0)|n, J, M_2\rangle = \sum_{m_\ell, m_s} C^{J, M_1}_{s, m_\ell, \ell, m_1}\langle n, \ell, m_\ell; s, m_1|\hat{\rho}(0)|n, \ell, m_\ell; s, m_2\rangle C^{J, M_2}_{s, m_\ell, \ell, m_2},
        \end{equation*}
        which, after using (\ref{eq:rho_0-compressed}), becomes
        \begin{multline*}
            \displaystyle\langle n, J, M_1|\hat{\rho}(0)|n, J, M_2\rangle = \sum_{m_\ell, m_s} C^{J, M_1}_{s, m_\ell, \ell, m_1}C^{J, M_2}_{s, m_\ell, \ell, m_2}\rho_{m_1, m_2}\times \\\langle n, \ell, m_\ell; s, m_1|n, \ell, m_\ell; s, m_1\rangle\langle n, \ell, m_\ell; s, m_2|n, \ell, m_\ell; s, m_2\rangle,
        \end{multline*}
        
        \noindent that, from the completeness relation, simplifies to 
        \begin{equation}
            \displaystyle\langle n, J, M_1|\hat{\rho}(0)|n, J, M_2\rangle = \sum_{m_\ell, m_s} C^{J, M_1}_{s, m_\ell, \ell, m_1}C^{J, M_2}_{s, m_\ell, \ell, m_2}\rho_{m_1, m_2}.
            \label{eq:rho_0-elements}
        \end{equation}
        
        We can rewrite equation \eqn{eq:rho_0-elements} in the following way
        \begin{equation}
            \hat{\rho}(0) = \sum_{m_\ell, m_s} C^{J, M_1}_{s, m_\ell, \ell, m_1}C^{J, M_2}_{s, m_\ell, \ell, m_2}\rho_{m_1, m_2}|M_1\rangle\langle M_2|
            \label{eq:rho_0-elements_op}
        \end{equation}
        
        It remains, then, to apply the time evolution operators in order to obtain the matrix elements of $\hat{\rho}(t)$. Using \eqn{activedef}, 
        \begin{equation}
            \displaystyle\hat{\rho}(t) = e^{iH\frac{t}{\hbar}}\left(\sum_{m_\ell, m_s} C^{J, M_1}_{s, m_\ell, \ell, m_1}C^{J, M_2}_{s, m_\ell, \ell, m_2}\rho_{m_1, m_2}|M_1\rangle\langle M_2|\right)e^{-iH\frac{t}{\hbar}},
            \label{eq:rho_t}
        \end{equation}
        This allows identifying how each component of the density matrix evolves in relation to the transformation induced by $H$. Then, the elements of the density matrix as a function of time will have the form
        \begin{equation}
            \langle n, J, M_1|\hat{\rho}(t)|n, J, M_2 \rangle = \langle M_1|e^{iH\frac{t}{\hbar}}\bigg(\sum_{m_\ell, m_s} C^{J, M_1}_{s, m_\ell, \ell, m_1}C^{J, M_2}_{s, m_\ell, \ell, m_2}\rho_{m_1, m_2}|M_1\rangle\langle M_2|\bigg)e^{-iH\frac{t}{\hbar}}|M_2\rangle
        \end{equation}
        and, therefore,  
        \begin{equation}
            \hat{\rho}(t) = \sum_{m_\ell, m_s} C^{J, M_1}_{s, m_\ell, \ell, m_1}C^{J, M_2}_{s, m_\ell, \ell, m_2}\rho_{m_1, m_2}e^{i(H_{M_1} - H_{M_2})\frac{t}{\hbar}}.
            \label{eq:active-density}
        \end{equation}

        The energy eigenvalues $H_{M_i}$ can be determined for a particle under rotation in a spherical box, which corresponds, with the potential used here, to a hydrogen-like atom under rotation such that it satisfies $|n, \ell, m_\ell\rangle = |n, \ell, m_\ell\rangle_H = |n, \ell, m_\ell; t = 0\rangle_S = \psi(r, \theta, \phi)$, which is given by
        \begin{equation}
            \displaystyle\psi(r, \theta, \phi) = A_{n, \ell, m}e^{-r/na_\kappa}\left(\frac{2r}{na_\kappa}\right)^\ell L^{2\ell + 1}_{n - \ell - 1}\left(\frac{2r}{na_\kappa}\right)Y^m_\ell (\theta, \phi), 
        \end{equation}
        where $A_{n, \ell, m}$ is the normalization constant and $a_\kappa$ the generalized Bohr radius, such that $\kappa \equiv |Q_p Q_N|$, with $Q_p$ and $Q_N$ the charges of the generalized particle and generalized nucleus, respectively. The energy eigenvalues are then given by
        \begin{equation}
            H_{M_i} = -\frac{\alpha^2 m}{2\hbar^2 n^2} - \hbar\omega_i M_i.
            \label{eq:eigenvalue}
        \end{equation}

        Replacing \eqn{eq:eigenvalue} in \eqn{eq:active-density}, we have
        \begin{gather*}
            \hat{\rho}(t) = \sum_{m_\ell, m_s} C^{J, M_1}_{s, m_\ell, \ell, m_1}C^{J, M_2}_{s, m_\ell, \ell, m_2}\rho_{m_1, m_2}e^{i\left(-\frac{\alpha^2 m}{2\hbar^2 n^2} - \hbar\omega_1 M_1 + \frac{\alpha^2 m}{2\hbar^2 n^2} + \hbar\omega_2 M_2\right)\frac{t}{\hbar}}\\
            \hat{\rho}(t) = \sum_{m_\ell, m_s} C^{J, M_1}_{s, m_\ell, \ell, m_1}C^{J, M_2}_{s, m_\ell, \ell, m_2}\rho_{m_1, m_2}e^{i\left(-\hbar\omega_1 M_1 + \hbar\omega_2 M_2\right)\frac{t}{\hbar}}        
        \end{gather*}

        With both states rotating with the same angular velocity in the $z$ direction, $\omega_1 = \omega_2 = \omega_z$,
        \begin{equation}
           \hat{\rho}(t) = \sum_{m_\ell, m_s} C^{J, M_1}_{s, m_\ell, \ell, m_1}C^{J, M_2}_{s, m_\ell, \ell, m_2}\rho_{m_1, m_2}e^{i\left(M_2 - M_1\right)\omega_z t}.
            \label{eq:active_rotation}
        \end{equation}

    \subsubsection{Passive rotation}
        In the reference frame of the particle, the detector rotates around the $z$-axis by an angle $\phi$ (via the action of the rotation operator $U(\phi, t) = D_z(\phi, t) = e^{-iJ_z\frac{\phi}{t}}$, where $J_z$ is the angular momentum operator in the $z$-direction). Following \eqn{activedef},
        \begin{gather}
            \hat{\rho}(\phi, t) = U^\dagger(\phi, t)\hat{\rho}(0, 0)U(\phi, t)\\
            \displaystyle\hat{\rho}(\phi, t) = e^{iJ_z\frac{\phi}{\hbar}}\hat{\rho}(0, 0)e^{-iJ_z\frac{\phi}{\hbar}}.
        \end{gather}

        Replacing $\hat{\rho}(0)$ defined in \eqn{eq:rho_0} and making $\phi = \omega_z t$,
        \begin{gather*}
            \displaystyle\hat{\rho}(t) = e^{i(J_z\omega_z)\frac{t}{\hbar}}\left(\rho_{m_1, m_2}|r\rangle\langle r'|\otimes |S\rangle\langle S'|\right)e^{-i(J_z\omega_z)\frac{t}{\hbar}}\\
            \displaystyle\hat{\rho}(t) = e^{i(J_z\omega_z)\frac{t}{\hbar}}\left(\sum_{m_\ell, m_s}C^{J, M_1}_{s, m_\ell, \ell, m_1}C^{J, M_2}_{s, m_\ell, \ell, m_2}\rho_{m_1, m_2}|M_1\rangle\langle M_2|\right)e^{-i(J_z\omega_z)\frac{t}{\hbar}}
        \end{gather*}
        and, therefore,
        \begin{equation}
            \displaystyle\hat{\rho}(t) = \sum_{m_\ell, m_s}C^{J, M_1}_{s, m_\ell, \ell, m_1}C^{J, M_2}_{s, m_\ell, \ell, m_2}\rho_{m_1, m_2}e^{i(J_{z, M_1}\omega_{z_1} - J_{z, M_2}\omega_{z_2})\frac{t}{\hbar}}.
            \label{eq:rho_passive}
        \end{equation}

       In this case, the eigenvalues of $J_{z, M_1}$ have the form $M_i\hbar$. Hence,
       \begin{equation*}
           \displaystyle\hat{\rho}(t) = \sum_{m_\ell, m_s}C^{J, M_1}_{s, m_\ell, \ell, m_1}C^{J, M_2}_{s, m_\ell, \ell, m_2}\rho_{m_1, m_2}e^{i(M_1\hbar\omega_{z_1} - M_2\hbar\omega_{z_2})\frac{t}{\hbar}},
       \end{equation*}
        where, from the same argument for the active rotation, $\omega_{z_1} = \omega_{z_2} = \omega_z$, that is,
        \begin{equation}
            \displaystyle\hat{\rho}(t) = \sum_{m_\ell, m_s}C^{J, M_1}_{s, m_\ell, \ell, m_1}C^{J, M_2}_{s, m_\ell, \ell, m_2}\rho_{m_1, m_2}e^{i(M_1 - M_2)\omega_zt}.
            \label{eq:passive_rotation}
        \end{equation}

        Comparing \eqn{eq:active_rotation} and \eqn{eq:passive_rotation}, the density matrix $\hat{\rho}(t)$ has the same form, except for the sign corresponding to the difference $M_2 - M_1 = -(M_1 - M_2)$ in the exponential term, which arises because the rotation is viewed from different reference frames. This indicates that the active and passive rotations of a generic spin particle under the action of a Coulomb potential are \textbf{equivalent}.

    \subsection{Position-varying magnetic field in two dimensions\label{magnetic}}

    \subsubsection{Active rotation}
        In the presence of a position-dependent magnetic field $\bm{B}(\bm{r})$,  the Hamiltonian of the system under active rotation acquires the contribution
        \begin{equation}
            \hat{H} = \frac{\hat{\bm{p}}^2}{2m} - \frac{\alpha}{r} - \bm{\omega}\cdot\hat{\bm{J}} - \hat{\bm{\mu}}\cdot\hat{\bm{B}}(\bm{r}) 
        \label{eq:magnetic_active_hamiltonian}
        \end{equation}
        where $\bm{B(r)} = \displaystyle\left(\Omega_1 + \frac{\Omega_2}{r} + \frac{\Omega_3}{r^2}\right)\hat{\bm{z}}$ and $\displaystyle\hat{\bm{\mu}} = \gamma\frac{q}{2m}\hat{\bm{S}}$  is the magnetic moment operator, with $\gamma$ being the gyromagnetic ratio and $q$ the particle's charge. 
    
        To find the energy eigenvalues of this system, the Hamiltonian is applied in the Schrödinger equation, resulting in
        \begin{equation}
            \displaystyle i\frac{\partial\psi}{\partial t} = \left[-\frac{1}{2m}\nabla^2 - \frac{\alpha}{r} - \omega_z \hat{J}_z - \gamma\frac{q}{2m}\hat{S}_z\left(\Omega_1 + \frac{\Omega_2}{r} + \frac{\Omega_3}{r^2}\right)\right]\psi. 
            \label{eq:magnetic_schrodinger}
        \end{equation}

        In spherical coordinates, considering only the radial and angular dependencies of the wave function, the radial component of the Schrödinger equation is given by
        \begin{gather}
            \frac{1}{r^2}\frac{1}{R(r)}\frac{d}{dr}\left(r^2\frac{dR(r)}{dr}\right) - 2m[V(r) - E] = 0\\
            \frac{1}{r^2}\frac{d}{dr}\left(r^2\frac{dR(r)}{dr}\right) - \textcolor{black}{2m}\left[\textcolor{black}{\frac{\ell(\ell + 1)}{r^2}} -\frac{\alpha}{r} -\omega_z M_j -\gamma\frac{q}{2m}M_j\left(\Omega_1 + \frac{\Omega_2}{r} + \frac{\Omega_3}{r^2}\right) - E\right]R(r) = 0
        \end{gather}

        Letting $\chi(r) = rR(r)$ and making the change of variable $x = 1/r$,
        \begin{multline}
            \frac{d^2\chi(x)}{dx^2} + \frac{2x}{x^2}\frac{d\chi(x)}{dx} +\frac{2m}{x^4}\bigg\{\omega_z M_j + \gamma\frac{q}{2m}M_j\Omega_1 + E +\left(\alpha + \gamma\frac{q}{2m}M_j\Omega_2\right)x + \\\left[\gamma\frac{q}{2m}M_j\Omega_3 - \textcolor{black}{\frac{\ell(\ell + 1)}{2m}}\right] x^2\bigg\}\chi(x) = 0,        
        \end{multline}
        which can be rewritten as
        \begin{equation}
            \frac{d^2\chi(x)}{dx^2} + \frac{2x}{x^2}\frac{d\chi(x)}{dx} +\frac{2m}{x^4}\left[- H_0 + H_1x + H_2x^2\right]\chi(x) = 0,
        \end{equation}
        where
        \begin{multline}
            H_0 = \textcolor{black}{-2m}\left(\omega_z M_j + \gamma\frac{q}{2m}M_j\Omega_1 + E\right) \eqcomma H_1 = \textcolor{black}{2m}\left(\alpha + \gamma\frac{q}{2m}M_j\Omega_2\right) \eqcomma \\H_2 = \textcolor{black}{2m}\left[\gamma\frac{q}{2m}M_j\Omega_3 - \textcolor{black}{\frac{\ell(\ell + 1)}{2m}}\right]. 
        \end{multline}

        By the Nikiforov-Uvarov method \cite{kayman, yazdankish2020}, the energy eigenvalues will be given by
        \begin{equation}
            \sqrt{H_0} = \frac{H_1}{1 + 2n \pm \sqrt{1 - 4H_2}},
        \end{equation}
        resulting in
        \begin{gather}
            \textcolor{black}{2m}\left(\omega_z M_j + \gamma\frac{q}{2m}M_j\Omega_1 + E\right) = \left[\frac{\textcolor{black}{2m}\left(\alpha + \gamma\frac{q}{2m}M_j\Omega_2\right)}{1 + 2n + \sqrt{1 - \textcolor{black}{2m}\left(2\gamma\frac{q}{m}M_j\Omega_3 - 2\textcolor{black}{\frac{\ell(\ell + 1)}{m}}\right)}}\right]^2\\
            E = 2m\left[\frac{\alpha + \gamma\frac{q}{2m}M_j\Omega_2}{1 + 2n + \sqrt{1 - 4\left(\gamma qM_j\Omega_3 - \textcolor{black}{\ell(\ell + 1)}\right)}}\right]^2 - M_j\left(\omega_z + \gamma\frac{q}{2m}\Omega_1\right)
        \end{gather}
    
        Thus, the density matrix operator given by \eqn{eq:active-density} for the active rotation will be given by
        \begin{multline}
            \hat{\rho}(t) = \sum_{m_\ell, m_s} C^{J, M_1}_{s, m_\ell, \ell, m_1}C^{J, M_2}_{s, m_\ell, \ell, m_2}\rho_{m_1, m_2}\exp\Bigg\{i\Bigg[2m\left(\frac{\alpha + \gamma\frac{q}{2m}M_1\Omega_2}{1 + 2n + \sqrt{1 - 4(\gamma qM_1\Omega_3 - \ell(\ell + 1))}}\right)^2 - \\ 2m\left(\frac{\alpha + \gamma\frac{q}{2m}M_2\Omega_2}{1 + 2n + \sqrt{1 - 4(\gamma qM_2\Omega_3 -     \ell(\ell + 1))}}\right)^2 + \left(M_2 - M_1\right)\left(\omega_z + \gamma\frac{q}{2m}\Omega_1\right)\Bigg]t\Bigg\}.
            \label{eq:magnetic_active_rotation}
        \end{multline}

        For this system, the states correspond to the wave function solution of the time-independent Schrödinger equation associated with \eqn{eq:magnetic_schrodinger}, which can be written as
        \begin{equation}
            \psi(r, \theta) = B_nr^{-\frac{H_1}{2\sqrt{H_0}}}e^{\sqrt{H_0}r}\left[-r^2\frac{d}{dr}\right]^n\left[r^{\frac{2n-H_1}{\sqrt{H_0}}}e^{-2\frac{\sqrt{H_0}}{r}}\right](-1)^mP_\ell^m(\cos\theta),
        \end{equation}
        where $B_n$ is the normalization constant and $P_\ell^m(\cos\theta)$ are the associated Legendre polynomials.

    \subsubsection{Passive rotation}
        In the system under passive rotation, the density operator takes the same form of \eqn{eq:passive_rotation}, derived in an equivalent way. Thus, for the rotations to be equivalent, the exponential term in \eqn{eq:magnetic_active_rotation} must be equal to $(M_1 - M_2)\Tilde{\omega}_z$. The issue is that, for this to occur in the passive rotation for this configuration, there must be an angular velocity $\Tilde{\omega}_z$ that depends on the quantum numbers $n$ and $m_s$, which is not possible since, in the rotating frame, $m_s$ is not a good quantum number. Therefore, for a space-varying magnetic field in the form defined here, the active and passive rotations of a generic spin particle \textbf{are not equivalent}.

    \subsection{Rotating cylindrical wells \label{cylinder}}
        Now, consider the case of a non-relativistic cylindrical well of radius $R$ and constant depth $U_0$, rotating with constant angular velocity $\omega$ about the $z$-axis, passing through its center of mass, as discussed in \cite{buzzegoli2023bound}. The Hamiltonian of this system reads
        \begin{equation}
            \hat{H} = \frac{\mathbf{\hat{p}}^2}{2m} - U_0 - \boldsymbol{\omega}\cdot\mathbf{\hat{J}},
            \label{eq:well_hamiltonian}
        \end{equation}
        and, if $R < 1/\Omega$ (the well is rotating slowly), the set of energy eigenvalues is
        \begin{equation}
            E_{Mnk_z} = -\frac{1}{2mR^2}y^2_{Mn} - \frac{M\lambda}{R} + \frac{k_z^2}{2m},
            \label{eq:slow_energy}
        \end{equation}
        where $y_{Mn}$, $n = 1, 2, ...$, is the set of solutions of the equation that satisfies the boundary conditions (see \cite{buzzegoli2023bound} for details), $\lambda = R\omega$, and $\kappa = \sqrt{2m\left(k_z^2/2m + |E + M\omega|\right)}$, $E + M\omega < 0$, with $k_z$ being the wave number.

        In a rapid rotation system $\left(R > 1/\omega\right)$, the set of energy eigenvalues is
        \begin{equation}
            E_{Mak_z} = \frac{1}{2m}\left(\omega^2x_{Ma}^2 + k_z^2\right) - U_0 - M\omega,
            \label{eq:rapid-energy}
        \end{equation}
        where $x_{Ma}$, $a = 1, 2, ...$, is the set of zeros of the M'th Bessel function.

        \subsubsection{Active rotation}
        For the slow rotation case, substituting \eqn{eq:slow_energy} in \eqn{eq:active-density},
        \begin{equation}
            \hat{\rho}(t) = \sum_{m_\ell, m_s} C^{J, M_1}_{s, m_\ell, \ell, m_1}C^{J, M_2}_{s, m_\ell, \ell, m_2}\rho_{m_1, m_2}\exp\bigg\{{i\left[\frac{1}{2mR^2}\left(y^2_{M_2n} - y^2_{M_1n}\right) + \frac{\lambda}{R}\left(M_2 - M_1\right)\right]\frac{t}{\hbar}}\bigg\},
        \end{equation}
        which can be rewritten as

        \begin{equation}
            \hat{\rho}(t) = \sum_{m_\ell, m_s} C^{J, M_1}_{s, m_\ell, \ell, m_1}C^{J, M_2}_{s, m_\ell, \ell, m_2}\rho_{m_1, m_2}\exp\left[\frac{i}{\hbar}\frac{1}{2mR^2}\left(y^2_{M_2n} - y^2_{M_1n}\right)t\right]\exp\left[\frac{i}{\hbar}\left(M_2 - M_1\right)\omega t\right].
            \label{eq:slow_active_density}
        \end{equation}

        Doing the same with \eqn{eq:rapid-energy}, the density matrix operator for the rapid rotation case is
        \begin{equation}
            \hat{\rho}(t) = \sum_{m_\ell, m_s} C^{J, M_1}_{s, m_\ell, \ell, m_1}C^{J, M_2}_{s, m_\ell, \ell, m_2}\rho_{m_1, m_2}\exp\bigg\{{-\left[\frac{1}{2m}\left(x^2_{M_1a} - x^2_{M_2a}\right) + \omega\left(M_2 - M_1\right)\right]\frac{t}{\hbar}}\bigg\},
        \end{equation}
        which can be rewritten as
        \begin{equation}
            \hat{\rho}(t) = \sum_{m_\ell, m_s} C^{J, M_1}_{s, m_\ell, \ell, m_1}C^{J, M_2}_{s, m_\ell, \ell, m_2}\rho_{m_1, m_2}\exp\left[\frac{i}{\hbar}\frac{1}{2m}\left(x^2_{M_1a} - x^2_{M_2a}\right)t\right]\exp\left[\frac{i}{\hbar}\left(M_2 - M_1\right)\omega t\right].
            \label{eq:rapid_active_density}
        \end{equation}

        \subsubsection{Passive rotation}
        Applying the rotation operator to both slow and rapid rotations would result in the same expression for $\hat{\rho}(t)$, since it depends only on $\hat{J}_z$, and not on the Hamiltonian eigenvalues. Hence, \eqn{eq:rho_passive} describes the passive rotation for both situations. Thereby, clearly \eqn{eq:slow_active_density} and \eqn{eq:rapid_active_density} are different from \eqn{eq:rho_passive}, indicating another case (a particle in a cylindrical well) in which the active and the passive rotations \textbf{are not equivalent}.

    \subsection{Coulombic potential in a 2D-rotating cylindrical well \label{coulcylinder}}
        In this section, we will evaluate one more time a Coulomb-like potential, but considering the cylindrical well previously discussed, following the same discussion in \cite{buzzegoli2023bound}. First, the wavefunction $\psi_0$, which is solution to a simpler stationary problem described by
        \begin{equation}
            \left(\bm{\nabla}^2 + \frac{2m\alpha}{r} -2m|E_0|\right)\psi_0 = 0,
        \end{equation}
        requires (from the fact that as $r \rightarrow \infty$, $\psi_0$ is finite) the quantization condition $n = n' - 1/2$, with $n' = 1, 2, ...$ and $|M| \leq n' - 1$. In a rotating system, $n'$ depends on $\omega$ and the energy levels are given by
        \begin{equation}
            E_{n', M} = -\frac{\alpha^2 m}{2\left[n'(\omega) -1/2\right]^2 }- M\omega.
            \label{eq:coulomb-rotating-well}
        \end{equation}

        These energy eigenvalues are completely different from \eqn{eq:eigenvalue} because of the dependence of $n$ on the magnetic quantum number in the quantization condition, which drastically changes the discussion of the equivalence between active and passive rotations in the presence of a potential proportional to $1/r$.

        \subsubsection{Active rotation}
            Using \eqn{eq:coulomb-rotating-well} in \eqn{eq:active-density}, we have
            \begin{multline}
                \hat{\rho}(t) = \sum_{m_\ell, m_s} C^{J, M_1}_{s, m_\ell, \ell, m_1}C^{J, M_2}_{s, m_\ell, \ell, m_2}\rho_{m_1, m_2}\times\\\exp\bigg\{i\alpha\left[\frac{1}{\left[n'_2(\omega) - 1/2\right]^2} - \frac{1}{\left[n'_1(\omega) - 1/2\right]^2 }\right]\frac{t}{\hbar}\bigg\}\exp\left[\frac{i}{\hbar}\left(M_2 - M_1\right)\omega t\right]
            \end{multline}
            
        \subsubsection{Passive rotation}
            From arguments already presented, it is clear that this system, in a passive rotation, would imply a density operator of the form \eqn{eq:passive_rotation}. Therefore, active and passive rotations for a particle under the influence of a Coulomb-like potential in a \textbf{cylindrical symmetry} are \textbf{not equivalent}. This is the opposite of the already shown Coulombc-like potential in a spherical symmetry. 
            
    \section{General principles: symmetries and equilibrium \label{symmetries}}
        The problems above are important as they illustrate an important difference between classical and quantum dynamics.  In the first case, cylindrical symmetry and spherical symmetry are equivalent {\em for rotations around the axis of the cylinder} (we'll call it $z$ here), because there is no fundamental uncertainty related to the commutators between generators.   
        
        In the quantum case, however, wavefunctions are always smeared in accordance to the uncertainty principle.   For spherical-type symmetries, this symmetry is bounded by the Wigner-Eckart theorem \cite{sakurai}, which makes the density matrix factorize into a part depending only on the Casimir and the rest.   
        \begin{equation}
            \label{wefact}
            \hat{\rho}=\sum_{J' M'} \langle J|\rho_0|J'\rangle\times \langle J,M|J',M'\rangle.
        \end{equation}
        
        For spatial symmetries, a representation of this is spherical harmonics, while for spin, one can use SU(2) matrices ($SO(3) \equiv SU(2)/Z_2$).   
        Cylindrical symmetry is described by the group $SO(2) \times T_1$, where $T_1$ are translations in the $z$ direction.  This symmetry group shares some, but not all generators with the spherical group, and consequently, the Casimir is different  ($J_{x}^{2}+J_{y}^{2}$).  In spacetime, its representation is given by Bessel functions.
        
        For a passive rotation, \eqn{wefact} will always be true as long as the system is rotationally invariant for that axis. But for an active one, it will depend on the symmetries of the potential causing the rotation, and if this potential breaks just one of the generators of the rotation group (e.g. if the potential has cylindrical rather than spherical symmetry), the factorization of \eqn{wefact} will break down. In this case, passive and active rotations will obviously not be equivalent, since, in one case, there is factorization and, in the the other, there is not.
        
        Thus, the equivalence of active and passive rotations 
        requires that {\em all} generators of the rotation group to commute, even if those are irrelevant to the physical rotation.  This is an inherently quantum effect, related intimately to the uncertainty (expressed via cumulants of operators) due to non-commuting operators included in the wavefunctions:   if a generator does not commute with the Hamiltonian, even one perpendicular to the rotation axis, the active rotation will introduce an uncertainty in the value of that operator, which will be absent in the passive rotation.   Since such uncertainties are measurable, active and passive rotations will not be the same.

        This reasoning has a parallel in the famous result of N. Bohr's doctoral research \cite{bohr} known as the Bohr–Van Leeuwen theorem. The vanishing of classical magnetization can be thought of as a consequence of the fact that the energy of the system is degenerate w.r.t. the magnetic field (which only acts on the angular momentum), and by the symmetries of spacetime, the partition function only depends on energy. In other words, since the system is invariant under rotations of the axis parallel to the magnetic field, but the only effect of the magnetic field is to induce time-dependent rotation along that axis of each particle, active and passive rotations coincide and, since classically passive rotations are irrelevant, so is the magnetic field.
        In the quantum regime, the presence of the magnetic potential in the momentum implies $[p_x,y] \ne 0$.  Even though, the system is only rotating in the $z$-axis and the wave-function structure is ``stretched" in the $x,y$ axis as well (the stretching direction is Gauge-dependent, but its extent is not).  This changes $p^{2}$ and, hence, $\ave{H}$.

        What happens if, in addition, the system is in global equilibrium?  In such case, the velocity of each energy cell can only be a killing vector \cite{buzzkill1,buzzkill2}. As examined in \cite{MSelch}, the equivalence of active and passive rotations trivially follows provided the underlying Hamiltonian is Lorentz invariant.
        The reason is that the maximally mixed state will become another maximally mixed state when translated by a Killing vector.  Thus, the second part of \eqn{criterion} is irrelevant and the first term will be satisfied by default by the values of temperature and polarizability, which satisfy the maximum entropy condition of \cite{buzzkill1,buzzkill2}. Instead of \eqn{wefact}, we have the killing vector condition and maximal mixedness in both passive and active rotations (in fact, the rigid body approximation is nothing else but the Killing vector criterion in the Galilean symmetry limit).
     
        Perfect local equilibrium \cite{son}, which implies ergodicity of the microscopic degrees of freedom in every infinitesimal cell  \cite{ergodic}, will give an analogous result provided the ergodic hypothesis is satisfied exactly for every cell, since the action on a maximally mixed state preserves maximal mixedness.   Away from these highly unrealistic limits, though, a violation of \eqn{criterion} will generally mean that active and passive rotations should be different.

        A similar discussion can be made for linearized general relativity, because of the principle of equivalence (combining full non-linear general relativity with quantum mechanics is, of course, an open problem, which we discuss briefly in the conclusion).
        
        In the Newtonian limit, the problem reduces to the one in \secref{coloumb}, where, as we saw, the passive and active rotations coincide.   For the relativistic linearized limit, as shown in \cite{feynman}, the tensorial structure of the theory, together with the Gauge consistency condition forces the interaction Lagrangian to be of the form 
        \begin{equation}
            \label{emtensor}
            L_{GR}=L_{matter}+T_{\mu \nu} h^{\mu \nu} \eqcomma T_{\mu \nu}  \sim  \left\{
            \begin{array}{cc}
                \partial_\mu \phi \partial_\nu \phi -\eta_{\mu \nu} \left( (\partial \phi)^{2}+m² \phi²\right)&\mathrm{scalar}\\
                \bar{\psi} i\left(\gamma^\mu \partial^\mu+\gamma^\nu \partial^\mu \right)+\frac{1}{2}m\eta^{\mu \nu}\bar{\psi}\psi &\mathrm{fermion}\\
                -F^{\mu \lambda} F^\nu_\lambda -\frac{1}{4}F^{2}+\frac{m^{2}}{2}\left(  \eta^{\mu \nu}A_\alpha A^\alpha - A^\mu A^\nu \right)&\mathrm{vector} 
            \end{array}
            \right.
        \end{equation}
        where $T_{\mu \nu}\sim  \frac{1}{\sqrt{\mathrm{det}\left(\eta_{\mu \nu}+h_{\mu \nu}\right)}} \frac{\delta L}{\delta h^{\mu \nu}}$ is the energy momentum tensor of the field, given by the right representation of the SU(2) group, and $h_{\mu \nu}$ the linear metric perturbation.  As shown in \cite{feynman}, it follows that the motion of a particle is a geodesic, with time-dilation given by factors of $\epsilon h_{\mu \nu} dx^\mu dx^\nu$, where $dx_\mu$ is a worldline element. Hence, 
        \begin{equation}
        S=\int d^{4} x L_{gr} \simeq \int (1+\epsilon) d t d^{3}x L_{matter}+\orderof{h^{2},\partial^{2}h}.
        \end{equation}
        For rotating motion (actually, for general motion in gravitational fields), this automatically satisfies \eqn{criterion}, since the generators of the rotation also generate the metric, giving rise to the rotation and time dilation appears in the ``active'' rotation (with $h_{\mu \nu}$ and the passive one (in the freely falling frame) in the same way).  
        Thus, for general relativity, active and passive motion in a gravitational field are locally equivalent, which is, in fact, one of the ways to see the equivalence principle.
        \section{Field theory \label{field}}
        The discussion in the previous section assumed ``static'' electric and magnetic fields in every frame, which is, at best, an approximation in the passive frame (the magnetic field in \eqn{eq:passive_rotation} will create a time-varying electric field, and indeed the ambiguity of which field is which to which observer was a motivator for relativity \cite{einstein}).   Deviating from this approximation will however inevitably result, at the same order in time gradients, in the creation of electromagnetic waves, forcing us to consider a field theory rather than a mechanical system if interactions are present \cite{necfields}.
        
        When one goes from quantum mechanics to quantum field theory, the infinity of degrees of freedom and the presence of inequivalent representations lead to the usual mathematical ambiguities.  Nevertheless, the definitions \eqn{activedef} and \eqn{passivedef} can be adapted straight-forwardly to a quantum field theory using the path integral method \cite{peskin}.    Since the active rotation is defined in the semi-classical limit, the background field method can be used to provide a suitable definition.  For electromagnetism, this means
        \begin{equation}
        \label{activedefqft}
        \hat{p}_0 \rightarrow \hat{p}_0 - eA_0 \eqcomma A_0 =V\hat{x}
        \end{equation}
        which can be straight-forwardly solved as it does not alter the Gaussian nature of the integral.

        The passive rotation has to be incorporated by a source term in the detector, 
        \begin{equation}
        \label{activedefqft}
        S=\int d^{4} x L(A,\psi) \rightarrow  \int d^{4} x \left[ L(A,\psi)+ J_\mu(x,t)A^\mu\right].
        \end{equation}
        Systematically working out the consequences of these is involved and has to be done on a case by case basis, but it is clear they continue, in general, to be different.  In the next few sections, we shall examine a few simple cases and try to generalize the implications.
    \subsection{The quantum fluctuation in an electromagnetic field of electrons in rings \label{bell}}     
    We are now ready to go beyond the test field assumptions of the previous section, and examine what happens when the passively rotating observer sees the static magnetic field of \eqn{eq:passive_rotation}.
    
        The Hamiltonian that takes into consideration the coupling between spin and angular rotation due the electron rotation in a ring is given by

        \begin{equation}
            \hat{H} = \frac{1}{2}\hbar \boldsymbol{\Omega}\cdot \boldsymbol{\hat{\sigma}},
        \end{equation}
        where we can write the angular rotation in the ring in two components: the first classical $\boldsymbol{\omega}$ and the second referent to quantum fluctuation $\delta \boldsymbol{\omega}$. Thus \cite{Bell},

        \begin{equation}
            \boldsymbol{\Omega} = \boldsymbol{\omega} + \delta \boldsymbol{\omega},
        \end{equation}
        where the quantum fluctuation is defined as \cite{Bell}:
        \begin{equation}
            \delta \omega_{\pm} = -\frac{e}{2mc}\left(g B'_{f\pm}+2E'_{fz}\right).
        \end{equation}
        
        $B'_{f\pm}$ and $E'_{fz}$ are magnetic and electric free fields, respectively. One possible interpretation is that these fields can be understood as the fields created due to the acceleration of the electron in the ring.

        The quantum fluctuation, $\delta \boldsymbol{\omega}$, occurs only in the active rotation, and there will be no influence of quantum fluctuation on the passive rotation. Thus, we will see that active and passive rotations differ in these scenarios. The possible observable is the electron polarization in both active and passive rotations.

        To understand this system, it is helpful to remember the famous ``Feynman's disk paradox''\cite[Ch.\ 17-4]{feynmanlectures}: a disk with static bead-like charges near its perimeter can freely rotate around its axis. The disk has a strong magnetic field parallel to its rotation axis, generated by a superconducting current. Initially, the system is at rest. The disk is then allowed to heat up until the magnetic field collapses, which generates a strong circular electric field tangent to the disk’s perimeter due to the presence of the charges. This would result in a net torque and, consequently, in the rotation of the disk. However, a paradox arises when considering that the angular momentum of the system must be conserved. Therefore, there should be no rotation, since the disk is initially at rest.

        The resolution of this paradox has to do with the angular momentum contained in the electromagnetic field, which of course, will, at infinity, contain radiated electromagnetic waves (a manifestation of the theorem \cite{necfields}).   Such asymptotic states will exist for active rotations, but not for passive ones (where they will appear as zero-momentum modes, as seen in \cite{higuchi}).   In the setup of \cite{Bell}, spin flips will be dynamically carried out by the emission and absorption of photons, which only happens when active rotations are performed.
    \subsection{Revisiting the bucket and general field theory lessons\label{qft}}
        We can now say something about quantum fields, relevant to the problems such as \cite{rot1,rot2,rot3,rot4,rot5,rot6,rot7} and calculations such as \cite{kim2023spin}.  Of course, the original problem of Newton is also a (classical) field theory problem, since water in a bucket is a classical field.
        Let us, therefore, say something about this problem in the language developed throughout this work.  The Hamiltonian for the system is
        \begin{equation}
            H= \frac{I \dot{\phi}_B}{2}+ \sum_{i=1}^N\frac{p_{i}^{2}}{2m}+ V_{contact}+V_{water}+V_{grav},
        \end{equation}
        where the first term is a simple rigid body rotating Hamiltonian, but the subsequent terms contain a ``large'' (call it $N$) number of degrees of freedom,
        \begin{equation}
            V_{contact} \propto \sum_{i=1}^N \delta\left(r_i-R \right)\left(\dot{\phi}_i-\dot{\phi}_b\right)^{2} \eqcomma V_{water} \sim \sum_{IJ=1}^N V(\vec{x}_i - \vec{x}_j) \eqcomma V_{grav} \sim \sum_{i=1}^N mg (\vec{x}_i)_z
        \end{equation}
        For $N \rightarrow \infty$ and no dissipation, $V_{water}$ converges to the Lagrangian of incompressible fluid dynamics, famously represented via volume preserving diffeomorphisms \cite{arnold}.  By inspection, the potential terms $V_{contact}$ and $V_{grav}$ and the generators of the diffeomorphism group of \cite{arnold} do not commute with $J_z$ (and more in general, the Galileo group), so we do not expect active and passive rotations to be the same even in the classical limit, thereby confirming \cite{newton}.

Let us now see what happens in quantum field theory.
The limit of QM from QFT is \cite{zee}
\begin{description}
    \item[Observable DoFs] are finite and put by hand as sources $J_{1,2,...}$
    \item[Potentials] are given by the expectation value of Wilson line operators between sources $W(J_1,J_2,...)$
\end{description}
Deriving potentials for generic fields, especially in the strongly coupled case, is of course a complicated task, but the sum rules in \cite{kim2023spin} give us some clue.    As argued in that work, one should follow the decomposition of \cite{ji} into a quark spin $S_q$ and a quark and gluon angular momenta $L_q,L_k$.  The first is local, the other two are not.   If a decomposition can be found of $W(J_1,J_2,...)$ into the relevant $S_q$,$L_{q,k}$ the latter will generally lead to differences in active and passive rotations.

The spin term is local and the other terms are not.  Thus, the only the first term should be invariant between passive and active rotations.    Constructing a Polyakov loop from $W(...)$ and making such an operator product expansion from the Polyakov loop should be sensitive to the difference between active and passive rotations of thermal media.

Of course, the identification of QFT correlators with quantum potentials only works for time-invariant problems.  Thus, quantum field theory can have ``imaginary'' values for the Wilson line operators, which signals an instability of the state in question \cite{peskin}.

Since intuitively only active rotations can cause a state to break \cite{charm1,charm2,charm3}, and bound states break via tidal forces, perhaps the imaginary part of $W(...)$ provides a QFT ``simple'' criterion that active and passive rotations are not in fact the same.     

An additional complication, with likely relevance to the ambiguities pointed out in \cite{sym1,sym2,sym3,sym4,sym5,sym6}, is that quantum field theory, because of the infinite number of degrees of freedom, does not meet the axioms of the Stone Von Neumann theorem, on which the symmetry arguments in \secref{symmetries} are based.   The fact that the vacuum  \cite{svaiter} does not, in general, have the symmetries of the Lagrangian is a direct consequence of this, and it can occur in several different ways, such as spontaneously broken symmetries (the breaking of the symmetries of the Lagrangian by infrared modes), anomalously broken symmetries (an interplay between infrared and ultraviolet modes via counter-terms) and boundary conditions (the so called Unruh and Hawking effects).
In setups such as \cite{sym1,sym2,sym3,sym4,sym5,sym6}, at least two of the effects numerated above are present, leading to ambiguity.  

This issue will be examined quantitatively in a future work.   We note that the equivalence of active and passive transformations inevitably implies that the effect of accelleration is indistinguishable from that of thermal baths, which some of \cite{sym1,sym2,sym3,sym4,sym5,sym6} disagree with.
On the other hand, because of  local operator non-commutativity, observing the system via an accellerated or rotating detector could in principle change it's microscopic structure.  We note however that the systems examined in \cite{sym1,sym2,sym3,sym4,sym5,sym6} have in a sense two temperatures, the internal one of the accellerating system and accelleration.   In such a case, given the arguments of \secref{symmetries} that equilibrium is what determines the relation between passive and active transformations, one must think carefully as to what is meant by equilibrium.  Considering a Gibbsian coarse-grained entropy rather than a Boltzmannian one for equilibrium (\cite{gauge} and ongoing work on Gibbsian hydrodynamics) might clarify whether active and passive non-inertial transformations coincide for a quantum field with multiple vacua.
    \section{Conclusions}
 This work has spanned a large range of topics, ranging from famous scientific texts \cite{newton,bohr} to technical problems of current interest in specialized topics such as QCD at finite temperature and vorticity \cite{rot1,rot2,rot3,rot4,rot5,rot6,rot7} (and more generally accelleration \cite{sym1,sym2,sym3,sym4,sym5,sym6}, where a controversy of the exact role of the ''Unruh effect'' in phase transitions exists in the literature).   As we have seen, the problem of whether passive and active rotations physically coincide in a given quantum system is, as expected, intimately related to the symmetries of the system and of the role symmetry has in quantum mechanics (which can be significantly different from the classical counterpart).   
 
 As we have shown both generally and in quantitative calculations, most ``realistic potentials'', not having a pure radial spherically symmetric dependence, will generally exhibit different behavior between active and passive limits.     The condition of equilibrium would also bring this difference to zero even for potentials that are not spherically symmetric, but such equilibrium has to be perfect, which is not a physically realistic assumption.
 
 While insights from quantum mechanics can be generalized to quantum field theory by symmetry arguments, ascertaining the quantitative discrepancy, necessary for phenomenology (is the difference between active and passive rotations a small correction or a large one in a given system?), is a much more challenging exercise for which at the moment we can not give even a qualitative recipe (beyond decompositions such as \cite{ji} and operator product expansions of the type used in \cite{kim2023spin,sasaki}.
 
While we concentrated on rotations, the arguments given here can be easily generalized to non-inertial frames.  A lot of the calculations done for this case have involved accellerations ``put in by hand''.  This has led to quite a lot of debates in the literature about whether the Unruh effect ``is real''.     Note that as shown in \cite{dunajski} the Unruh effect has a non-relativistic remnant, making it amenable to the sort of quantum mechanical analysis performed in this work.
Mixed particle states, allowed in quantum field theory, yield additional issues, particularly if the mixing is between mass and charge states (as in neutrinoes) rather than between two types of charges (as in quarks):  As shown in \cite{dharam1,dharam2} a vacuum of mass states will yield an automatic violation of passive vs active rotations due to the thermal vacuum being defined in terms of mass states.   The alternative, assuming the vacuum to be charged states, will yield a small violation of Lorentz symmetry and a non-thermal Unruh spectrum \cite{blas1,blas2}.

Understanding general criteria for the differences in passive and active accellerations (the latter with an explicit interaction with a classical field included \cite{elec1,elec2,elec3}) could be central to this question.   In this respect, a clear example of this is provided by the recent extension of the COW experiment \cite{cow} to freely falling wavepackets \cite{fallingcow}: Looking at FIg. 1 of \cite{fallingcow} one can think of an ''active'' transformation as a detector comoving with the red wavepacket measuring the phase shift of the blue wavepacket, and a ''passive'' transformation as the opposite' (the detector is assumed to be performing measurements in intervals longer than the width of the wavepacket).  The equality of the two phase shifts is a good illustration that in gravitational dynamics active and passive transformations are indeed the same.   

 Hence, beyond heavy ion physics, the sort of questions examined in this paper are of relevance for the rapidly growing field of quantum reference frames \cite{refe1,refe2,refe3} and their relationship to the quantization of gravity.
 Ultimately, the arguments around Newton's bucket were satisfactorily settled by Einstein, who worked out that it is the local gravitational field (produced by the bucket, the water, the earth etc.) that determines the family of local ``inertial'' coordinate systems which determine which set of transformations are equivalent between active and passive.    Of course this is still the last theory that is resistent to quantization.   Perhaps, the theoretical \cite{refe1,refe2} and experimental \cite{cow,werner} study of quantum systems in non-inertial reference frames could give clues \cite{medeco1,medeco2} to the right procedure to accomplish this task.

 This work would not have been written without G.T.'s two visits to Yonsei university made possible by Bolsa de pesquisa FAPESP 2023/06278-2 (a joint program with the National Research foundation of Korea) and the hospitality of Su Houng Lee and his group.  The problems treated here arose during the discussions with Su Houng Lee and Hyung Joo Kim, who also provided critical discussions and feedback on which the arguments made here crucially depend. 
 G.T. also acknowledges support from Bolsa de produtividade CNPQ 305731/2023-8, . K.J.G. is supported by CAPES doctoral
fellowship 88887.464061/2019-00
    \bibliographystyle{apsrev4-1}
    \bibliography{ref}

\end{document}